\documentclass[preprint2]{aastex}
\def\Msun{\>{\rm M_{\odot}}}

\def\kmsec{km s$^{-1}$~}

\newcommand{\nsns}{NS-NS~}
\newcommand{\bhns}{BH-NS~}

\newcommand{\bhbh}{BH-BH~}
\newcommand{\bhpsr}{BH-PSR~}
\begin{document}

\title{Nova Sco and coalescing low mass black hole binaries\\
as LIGO sources}

\author{Michael S. Sipior\altaffilmark{1}}
\affil{Department of Astronomy \& Astrophysics, 525 Davey Laboratory, \\
Pennsylvania State University, University Park, Pa 16802}

\author{Steinn Sigurdsson\altaffilmark{2}}
\affil{Department of Astronomy \& Astrophysics, 525 Davey Laboratory, \\
Pennsylvania State University, University Park, Pa 16802}

\altaffiltext{1}{sipior@astro.psu.edu}
\altaffiltext{2}{steinn@astro.psu.edu}

%\clearpage

%%%%%%%%%%%%%%%
% Abstract
%%%%%%%%%%%%%%%

\begin{abstract}

Double neutron star (\nsns) binaries, analogous to the well known Hulse--Taylor
pulsar PSR~1913+16 \citep{joe1}, are guaranteed-to-exist sources
of high frequency gravitational radiation detectable by LIGO.
There is considerable uncertainty in the estimated rate of
coalescence of such systems \citep{esp1,nar1,vik3},
with conservative estimates of $\sim 1 $ per million years per galaxy,
and optimistic theoretical estimates one or more magnitude larger.
Formation rates of low-mass black hole-neutron star binaries
may be higher than those of
\nsns binaries, and may dominate the detectable LIGO signal rate.
Rate estimates for such binaries are plagued by severe model uncertainties.
Recent estimates suggest that \bhbh binaries do not coalesce at significant
rates despite being formed at high rates \citep{zwart1,don98}.

We estimate the enhanced coalescence rate for \bhbh binaries due to weak
asymmetric kicks during the formation of low mass black holes like Nova Sco
\citep{brandt1}, and find they may contribute significantly to the LIGO signal
rate, possibly dominating the phase I detectable signals if the range of BH
masses for which there is significant kick is broad enough. For a standard
Salpeter IMF, assuming mild natal kicks, we project that the $R_6$ merger rate
(the rate of mergers per million years in a Milky Way-like galaxy)
of \bhbh systems is $\sim0.5$, smaller than that of \nsns systems.
However, the higher chirp mass of these systems produces a signal nearly
four times greater, on average, with a commensurate increase in search
volume. Hence, our claim that \bhbh mergers (and, to a lesser extent, \bhns
coalescence) should comprise a significant fraction of the signal seen by
LIGO.

The \bhbh coalescence channel considered here also predicts that a substantial
fraction of \bhbh systems should have at least one component with near-maximal
spin ($a/M \sim 1$). This is from the spin-up provided by the fallback
material after a supernova. If no mass transfer occurs between the two
supernovae, both components could be spinning rapidly. The waveforms produced
by the coalescence of such a system should produce a clear spin signature, so
this hypothesis could be directly tested by LIGO.

\end{abstract}

%%%%%%%%%%%%%%%
% Keywords
%%%%%%%%%%%%%%%


%%%%%%%%%%%%%%%
% Beginning of main text
%%%%%%%%%%%%%%%

%\clearpage

\section{Introduction}

With the arrival of LIGO, and other planned gravitational radiation
observatories, the nascent field of gravitational radiation astronomy is
set to prosper, if there are detectable sources in the local universe.
A considerable amount of effort has been directed towards identifying
potential gravitational radiation sources, and the relative rate of
contribution of these sources to the anticipated signals. The canonical
scenario envisioned is the final coalescence of a binary neutron star
system (NS-NS), where the pair's orbital energy has been radiated away by
gravitational waves.

\nsns merger \emph{can} produce a copious signal for LIGO, if they occur
at high enough a rate locally --- current estimates for LIGO phase I place
the maximum detection radius for such an event at $\sim\!20\!$~Mpc or
less \citep{vik3}. The merger rate of \nsns binaries can be estimated
from observed systems \citep{esp1,nar1}, such estimates are plagued by
small number statistics and possible observer biases \citep{vik1,vik3}.
Alternatively, the rate can be estimated from ab initio theoretical models
(see review by \citealt{gri01}, also \citealt{zwart1,bel:2001,fryer:1999,
bloom1999}).

A number of authors have explored binary population synthesis models, making
a number of assumptions about the input physics, leading to rates consistent
with observational constraints, but uncertain by 1--2 orders of magnitude
\citep{zwart1,vik1,vik2,bloom1999,don98,bel1,lip1,bra1,bel:2001,fryer:1999}. Conventionally, the
rates are expressed in terms of $R_6$, the merger rate per million years per
Milky Way-like galaxy, assuming normal rates of star formation. Then the
rate in the local universe is the integrated rate over the number density of
galaxies, the rate per galaxy, scaled to the Milky Way rate, per detection
volume. \citet{vik3} find a LIGO I event rate of $3\times 10^{-4}\ {\rm
y^{-1}}$ for NS-NS mergers within 20 Mpc, assuming $R_6=1$. For a given
detector sensitivity, higher mass coalescences are detectable to larger
volumes, with the detection distance, $d_L \propto M_{chirp}^{5/6}$, where
$M_{chirp} = (m_1m_2)^{3/5}/(m_1+m_2)^{1/5}$ \citep{tho94}. Since black
holes are expected to have masses several times larger than neutron stars,
and the event rate scales as $d_L^3$ (for $d_L$ small compared to the size
of the universe), a black hole coalescence rate of order $R_6$ implies event
rates 2--3 orders of magnitude higher.

The problem of accurately estimating the coalescence rate of compact
binaries can be appreciated by noting that the type II supernovae rate
in the Milky Way is 1--2 per century, implying that the rate of type II
supernova in binaries is about $10^{-2} \ {\rm y^{-1}}$, given a 50\%
binarity rate. Estimates of the coalescence rate are canonically close
to $R_6 = 10^{-6} \ {\rm y^{-1}}$ per Milky Way. So the branching ratio
for type II supernovae to form merging systems is of the order $10^{-4}$.
Calculating the mean rate, averaged over all scenarios for coalescence,
in the local universe, is hard, with small errors of assumption about the
physics of stellar evolution leading to large fractional changes in the
branching ratios of particular channels for mergers.

Theoretical models require a series of assumptions, about the mass function
of high mass stars, the cut-off points for which zero age masses lead to
NS or BH formation, the binary fraction and mass ratio distribution, the
amount of mass loss during (binary) stellar evolution, and the amplitude
and distribution of natal kicks. Secondary assumptions, that are usually
not explored in detail, include the dependence of all of the above on
metallicity and environment in which the massive stars form; and the
possibility that the NS/BH cut-off, and the natal kick, may depend on
stellar rotation and magnetic fields. In fact, it is not implausible that
the NS/BH formation boundary is ``fuzzy'', that there is not a sharp border
in zero-age mass between stars that form neutron stars and those which form
black holes. A further potential confounding effect, is that natal kicks
are generally assumed to be random, but may in principle be correlated with
some macroscopic property of the progenitor star, such as rotation (e.g.
\citealt{spr98,pfahl:2001}).

There are two arguments against a high rate of coalescence of compact
binaries: most high mass binaries become unbound during the near instantaneous
mass-loss at supernova; the second problem is that of the `kicks' believed
to be inflicted upon the NS from an asymmetrical supernova blast. While
the range of energies imparted by the kick is a subject of intense debate
\citep{lyne:1994,han97,tauris:1999,cordes:1998,fryer:1998,arz01}, the
resulting velocity change is believed to be of the order $250~\mbox{km
sec}^{-1}$ for a $1.4~\!\mbox{M}_{\odot}$ neutron star. A properly aligned
kick can allow the system to remain bound, albeit with dramatically altered
orbital parameters, but a random kick is unlikely to be directed so
fortuitously, with the extra energy making a bound final state even less
likely.

An alternative to the \nsns scenario is one in which both stars
are of sufficient mass to end up as black holes. The minimum mass
required for black hole formation is not well known, and is heavily
influenced by the evolutionary history of the black hole candidate (see
\citealt{fryer1999,fryer2001}). Accretion induced spin-up of the star prior
to collapse could result in a higher minimum mass threshold, while the
presence of strong winds could drop the progenitor mass below the required
minimum. Finally, unless an unusually flat IMF is assumed (as apparently
seen in certain starburst galaxies, see, for example, \citealt{doane1993}),
few of these objects will be produced in the first place.

Nevertheless, precursors to black hole binary systems are more likely to
survive both supernovae due to the smaller fractional mass loss during the
event. Depending upon the mass of the star, a significant amount of material
can fall back upon it just after the blast. Indeed, a sufficiently massive
progenitor may not undergo a supernova at all, reaching a point where nearly
all of the material is quickly reabsorbed \citep{fryer1999}. After losing a
significant fraction of mass to stellar winds and (possibly) two supernovae,
a typical bound binary system will have expanded considerably, with most
$P_{orb} > 10\!$ days. Long-period systems do not decay in less than a
Hubble time from gravitational radiation, and hence will never reach a
frequency range useful to LIGO. However, including asymmetric supernova
kicks changes the picture significantly. Assuming that the total momentum
change is approximately the same for all supernovae, the velocity imparted
to the nascent black hole will be scaled to its mass, via $\Delta v_{BH} =
\Delta v_{NS} M_{NS}/M_{BH}$ (see Grishchuk et al 2001 for review). For a
typical $7\,\mbox{M}_{\odot}$ hole, this gives a kick of the order $\sim 50
\ {\rm km\, s^{-1}}$, comparable to or larger than the orbital speed, but
not so large as to always rip the system apart. Systems that remain bound
are generally in highly eccentric orbits, which expedites their merger
through emission of gravitational radiation.

Ultimately, the point is this: the trace of the metric perturbation from
gravitational waves, varies as $h \propto M_{chirp}^{5/6}$. As the maximum
detection radius scales linearly with $h$, the maximum search volume scales
as $V \propto M_{chirp}^{5/2}$. The implied several hundred-fold increase in
search volume means that, even if only a small fraction of these systems can
merge in $\tau < 1/H_0$, they may dominate the detected signal.

Our approach is to concentrate on estimating the event rate from one
particular coalescence channel. The total branching ratio for coalescence is
of course the sum of all possible channels, and it is possible that other
channels contribute significantly to the event rate, possibly dominating the
total rate; for example, dynamical evolution of cluster binaries may create
new channels for merger with high coalescence rates \citep{sig1,zwart2},
or accretion induced collapse of neutron stars with soft cores may lead to
enhanced rates \citep{bethe1}. Here, rather than trying to estimate the
total event rate, we make an observationally motivated estimate of the rate
for one particular channel.

\section{Example of Nova Sco}

Nova Sco 1994 (GRO J1655-40) is a strong candidate for being a black hole
binary \citep{bai95,oro97}, with the primary being a $6.3 \pm 0.5 \Msun $
black hole \citep{gre01,sha99}. Perhaps its most remarkable feature is an
unusually high space velocity, whose lower limit of 106 \kmsec was, until
recently, several times greater than any other known black hole transient
\citep{sha99,brandt1}. The likely true space velocity is greater by a factor
of $\sqrt{3}$, adjusting for mean projection effects.

A number of scenarios have been put forward to explain the unusual speed
of Nova Sco. \citet{brandt1} point out that the momentum component of
Nova Sco along the line of sight is comparable to that of a single
$1.4\,\mbox{M}_{\odot}$ neutron star, having received a natal kick in the
range 300--700 \kmsec. This is not an unreasonable value for a neutron star
kick \citep{lyne:1994,cordes:1998,fryer:1998}, and so lends strength to the
possibility that Nova Sco can be explained by the primary experiencing a natal
kick prior to formation of the black hole. Invoking Blaauw-Boersma kicks
\citep{blaauw} as the sole acceleration mechanism invites difficulty, as the
low mass of the secondary means that most Blaauw-Boersma kicks strong enough
to give the observed speed would also disrupt the binary system completely
\citep{brandt1}. This is not to say that such a scenario is impossible (see,
for example, \citealp{nel99}). However, the space velocity measurement of
the X-ray nova XTE J1118+480 has provided a further example of a black hole
binary with a high-velocity Galactic-halo orbit. \citet{mirabel:2001} used the
VLBA to obtain a precise proper motion for the system, calculating a speed of
145$\pm 35$ \kmsec, with respect to the local standard of rest. The system
primary has a mass function of 6.0$\pm 0.4$M$_{\odot}$ \citep{mcclint:2001},
with a faint ($\sim19\,$mag) optical counterpart. To accelerate this system to
the observed peculiar velocity using Blaauw-Boersma kicks alone, roughly 40
M$_{\odot}$ of material would have to be expelled during stellar collapse, an
implausibly large amount of matter. This issue remains contentious, however, as
calculations in \citet{nel99} show that the velocity of Nova Sco, at
least, might be explained by a symmetric supernova where the black hole
progenitor lost more than half of its mass. This is marginally possible, but
becomes much more tenuous if Nova Sco turns out to have a significant
transverse velocity.

One principal difficulty with the natal kick premise is arranging for
sufficient neutrinos to drive a supernova explosion prior to being trapped
by the formation of an event horizon, ejecting neutrinos or other material
asymmetrically. Few neutrinos escape if the horizon
forms over the dynamical time of the collapsing core \citep{gour:1993}. The
drop in the resulting neutrino heating of the envelope allows most or all of
the envelope to fall back onto the nascent black hole (see, for example, the
hydrodynamical simulations of \citealt{janka:1996}), hence preventing the
supernova. However, there is now strong evidence that a supernova must have
taken place in the Nova Sco system, from estimates of metal abundances in
the atmosphere of the secondary star. \citet{isr99} show that the secondary,
an F3--F8 IV/III star, possesses a dramatic enhancement (factors of six to
ten) in $\alpha$ elements such as oxygen, magnesium, silicon and sulfur
(and see also the discussion in \citealt{pods:2001}).
Interestingly, no significant enhancement in iron was found. The implication
is that the progenitor of the black hole did experience a supernova during its
formation, and a significant amount of the former atmosphere was captured by
its companion. As most of the iron core went on to collapse to a singularity,
no enhancement of this particular element is seen. The question is then how
long after the supernova it took for the black hole to form. One possibility
is that the star, at the instant of the supernova, experienced sufficient
rotational support to avoid collapse for an extended period of time, until
spin-down allowed the system to collapse to a black hole some time later.
Another possible scenario has some of the material cast off in the supernova
explosion recaptured by the newly-formed neutron star, elevating its mass
above the threshold for collapse. In either case, a supernova is seen, with
a corresponding asymmetric natal kick. More massive progenitors would form a
horizon directly, with no possibility for a kick. We discuss limits for this
behaviour in the next section. Nova Sco 1994 and XTE J1118+480 provide strong
evidence that black holes do, in fact, experience natal asymmetric kicks, at
least under some conditions. The resulting orbital eccentricities tend to
accelerate binary coalescence from the enhanced radiation emitted at every
periastron passage.

\section{Population synthesis}

We make use of a binary evolution code developed by \citep{pols1994}, and
modified for use in \nsns systems by \citep{bloom1999}. Our extension of the
code allows for evolution to the black hole state, with assumptions about
the mass function of such objects at the time of collapse.

Initially, the code chooses the mass of the primary from a given mass
function. For this work, we have chosen two power law IMFs, with indices
$\alpha$ of $-2.0$ and $-2.35$ (the latter, of course, being the Salpeter
IMF). In both cases, we established a lower cutoff of $4\!$~M$_{\odot}$,
confining the code to an interesting range of initial masses; i.~e., where at
least one supernova is possible. Our primary stellar models are taken from
\citet{maeder:1989}. The helium star models used are a mix of models from
\citet{habets:1986} and \citet{pacz:1971}. We assume that the mass ratio
distribution between the two components is flat. After choosing an initial
separation $a$ and eccentricity $e$ after \citet{pols1994}, the code evolves
each binary system until both components have reached their final degenerate
form, accounting for mass-transfer-induced stellar regeneration and stellar
winds. During the common-envelope phase, the orbit is circularised, and the
orbital energy is reduced by the binding energy of the envelope divided by
the common-envelope efficiency parameter, which we take to be 0.5. In other
words, the orbital energy is reduced by twice the envelope binding energy.
Neutron stars are formed from progenitors with ZAMS masses of between 8 and
20 M$_{\odot}$, inclusive, and are always given a mass of 1.4 $M_{\odot}$.
More massive stars end up as black holes. As noted in the \emph{Introduction},
this boundary is likely to be ``fuzzy'', i.e., not a monotonic function of the
progenitor mass, as it is strongly coupled to the spin state of the star prior
to collapse, a quantity which our code simply does not track. Even assuming
this was known to perfect accuracy, it is far from trivial to estimate the
effects of magnetic field and rotational support vis-a-vis the compact
object's end state.

The black hole mass function (i.~e., the post-collapse mass of a BH, given
its mass just prior to the explosion) is highly speculative at this point,
and is almost certainly not merely a function of initial mass, but also
of angular momentum, to the extent that this determines the fraction of
material falling back onto the collapsing star. In order to experience
a kick, the black holes formation must be delayed somewhat, either from
rotational support, or because event horizon formation occurs only after
delayed fallback of mass initially ejected from the core. \citet{fryer1999}
has performed core-collapse simulations in order to explore the critical
mass for black hole formation, and their final masses. As a best working
guess, we have constructed a mass relation from a quadratic fit to the
limited data set found in \citet{fryer1999}. The mass of the black hole
at formation ($M_{BH}$) is related to the ZAMS mass of the progenitor
($M_0$) by $M_{BH} = \left(M_0/25\right)^2 \times 5.2 M_{\odot}$. We have
also chosen a simple criterion for whether a black hole will receive an
asymmetric neutrino kick during collapse; namely, all objects below 40
M$_{\odot}$ (referring to the ZAMS mass) experience a random kick. Above
this limit, objects collapse directly to a black hole, with no kick. We
explore the effect of ignoring black hole kicks in the next section.

As a last step, the results of the code are normalised to the supernova rate of
the Milky Way, $0.01\;\mbox{yr}^{-1}$. Here, the code produces some arbitrarily
large number of complete evolutionary sequences, including a large number of
supernova events. Scaling all these events to the Milky Way SNR allows us to 
generate event rates out of data not previously ordered in time. This
effectively simulates a constant star-formation rate, where the population of
merger candidates has reached a dynamic equilibrium.

\section{Results}

Tables \ref{r6rates} and \ref{r6nobhkicks} summarize the investigation.
For each IMF, six simulations (generating $10^6$ binary systems each) were
run, varying two parameters: $K_{\mbox{max}}$, the upper limit of the
imparted kick speed (acting on a $1.4\!$ M$_{\odot}$ NS), and $\sigma$, the
dispersion in the Maxwellian distribution of the kick speed. Maximum kick
values were chosen at 500 \kmsec (for $\sigma$ = 90 \kmsec and $\sigma$ =
190 \kmsec), and 1000 \kmsec (for $\sigma$ = 450 \kmsec). The inclusion of
the 90 \kmsec kick intensity is based upon recent observational work on the
velocity distribution of pulsars by \citet{pfahl:2001}, which suggests a
bi-modal kick distribution centered around $\sim100$ and $\sim500$ \kmsec.
Further discussions concerning this bi-modality can be found in
\citet{arz01,cordes:1998,fryer:1998}.
This range has been touched on in previous work \citep{gri01}, but we have
looked at it specifically as a new observationally-motivated choice in kick
magnitude. The following columns show the $R_6$ rate of \bhbh, \bhns, \nsns
and \bhpsr pairs generated in the sample. $R_6$ is a rate of 1 event per Myr
per Milky Way galaxy, found by scaling event rates to the estimated local
supernova rate of 0.01 yr$^{-1}$. One must take into account that not all
stars (and hence not all supernovae) occur in binaries; hence a binarity
fraction must be assumed, and we take that value to be 0.5 throughout (i.~e.
roughly two-thirds of supernovae occur in a binary system). The alternating
columns show the $R_6$ of such pairs expected to coalesce through emission
of gravitational radiation in less than the Hubble time (here, taken as
10~Gyr). Table \ref{r6rates} details the formation and merger rates when
black holes and neutron stars are given kicks as described above. Table
\ref{r6nobhkicks} gives the relevant rates when black holes are not allowed
to have asymmetric kicks of any kind (even if below 40 M$_{\odot}$).

\begin{figure}
\epsscale{1.0}
\plottwo{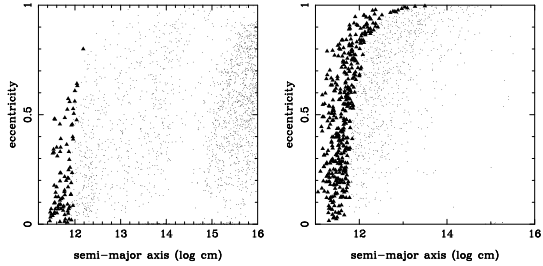}{f1b.ps}
%\epsfxsize=5.0cm
%\epsfysize=5.0cm
%\epsfbox{fig.ps}
\figcaption{\label{avse}A plot of the semi-major axis versus eccentricity for 
all \bhbh (left) and \bhns (right) systems generated in a single run of
$10^6$ binaries. Triangles designate systems which will coalesce in less
than 10 Gyr, whilst circles are systems stable against gravitational
coalescence. For the \bhbh systems, a clear distinction between merging and
stable binaries is evident. The bulk of those \bhns binaries
which remain bound after the second supernova also merge within a Hubble time.
The IMF index used is $\alpha = 2.35$, the Salpeter IMF. The
dispersion in the Maxwellian kick applied at each supernova is $\sigma = 90$
\kmsec, with the kick maximum $K_{max} = 500$ \kmsec. All kick
speeds apply to a 1.4 M$_{\odot}$ neutron star; the actual $\Delta v$ is found
by scaling the kick linearly with the compact object mass.}
\end{figure}

\begin{figure}
%\epsscale{0.5}
\plotone{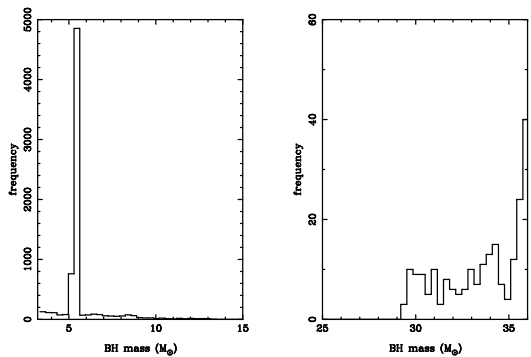}
\figcaption{\label{bhbhhist}Two histograms displaying mass distributions for
\bhbh pairs that remain bound after the second supernova. The parameters of the
run shown are an IMF parameter $\alpha = 2.35$, kick parameters $\sigma = 90$
\kmsec and $K_{max} = 500$ \kmsec. The left histogram shows that black hole
masses clump in two locations; one around 5 M$_{\odot}$, and the other 
at 35 M$_{\odot}$. We associate the lower-mass BHs with 20--40 M$_{\odot}$ ZAMS
progenitors, which lose much of their mass through a supernova (and hence
experience a natal kick). The high-mass set came from more massive progenitors,
and collapsed to a black hole without a supernova or the resulting kick.}
\end{figure}

\begin{figure}
\epsscale{0.5}
\plotone{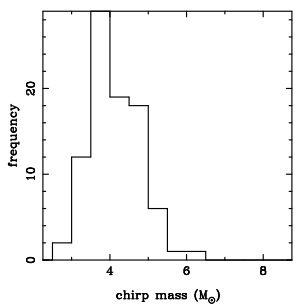}
\figcaption{\label{chirp}Shows the distribution of chirp masses for all
\bhbh systems that will eventually merge. The gravitational wave amplitude
scales as $M_{chirp}^{5/6}$, so this parallels the distribution of signal
strengths one could expect from a \bhbh population.}
\end{figure}

Figure \ref{avse} show examples of the generated data sets. Here, the
separations of the \bhbh and \bhns systems after the second supernova (i.~e.,
stellar evolution has ended) are plotted against the post-mortem orbital
eccentricity. Dots represent systems stable against orbital decay, whilst
triangles represent those systems which will undergo coalescence in less
than a Hubble time. Figure \ref{bhbhhist} shows the distribution of masses
in surviving \bhbh binaries, irrespective of whether the system is unstable to
orbital decay.. A clear bifurcation in mass is seen, with the
vast majority at around 5--6 M$_{\odot}$, and another, smaller concentration
at roughly 35 M$_{\odot}$. The low end is comprised of black holes with
20--40 M$_{\odot}$ ZAMS progenitors, which lose much of their mass through
a supernova (and hence experience a natal kick). The higher mass group are
from more massive progenitors, and collapsed to a black hole without a
supernova or the resulting kick. The histogram in figure \ref{chirp} shows the
''chirp'' mass distribution for all \bhbh systems that eventually merge from
gravitational radiation. As the wave amplitude scales like $M_{chirp}^{5/6}$,
this plot, coupled with the sensitivity of the LIGO detector, allows us to
estimate a detection rate for \bhbh binaries. By way of comparison, two 1.4
$M_{\odot}$ objects produce a chirp mass of 1.22 $M_{\odot}$.

As can be seen from column 8 in tables \ref{r6rates} and \ref{r6nobhkicks}, 
most \nsns pairs that remain intact coalesce in $\tau
< 1/H_0$, irrespective of IMF and kick strength. This is perhaps not
surprising, since \nsns systems must generally be tightly bound to remain
intact through the second supernova. For \nsns systems, we estimate
the Galactic merger rate to range from $\sim$0.04--8 Myr$^{-1}$, with
a strong dependence on the severity of the natal kick. Given a likely
LIGO I detection radius of $\sim20$~Mpc for such systems, this would
imply a conservative detection rate of 0.01--3 $\times10^{-3}$yr$^{-1}$
\citep{esp1,vik2}, with a likely mean rate of $\sim$10$^{-3}$ yr$^{-1}$.
These estimates are an order of magnitude lower than those stated in
\citet{gri01}.

A more unexpected result was the relative frequency of \bhns pairs relative
to bound NS systems. While far fewer \bhns systems are formed, their mass
allows them to weather two supernovae more frequently than two neutron
stars. From an estimate of the \bhns formation rate, we can say something
about the expected frequency of \bhpsr systems. For purposes of discussion,
we divide these systems into three categories, based on formation mechanism.
The first type results from the standard scenario: a massive black hole
forms first, with a regular short-lived pulsar at a later epoch. The second
possibility is that the black hole progenitor transfers a substantial
amount of matter on to the neutron star progenitor early in the evolution
of the system. This results in the neutron star forming \emph{first}, and
allows the black hole progenitor to continue to spill matter through its Roche
lobe on to the slowing pulsar, thereby recycling it. The last possibility 
is that the system
from scenario two is disrupted by the black hole formation, resulting in
a (possibly recycled) pulsar and single black hole. This channel is not
relevant for our purposes, as it ceases to be a potential LIGO source.
The $R_6$ rates for the remaining formation channels are shown in tables
\ref{r6rates} and \ref{r6nobhkicks}. The \bhns column covers all formation
channels for a \bhns system, whereas the BH--PSR$_2$ column considers only
that subset of \bhns systems where a recycled pulsar is formed first, via
channel two, as described above. Surprisingly, of the systems that go on to
form a \bhns pair, around 40\% experience the reversal described in case two
above, where the initially more massive star forms a compact object last. Only
0.5\% of the \bhbh progenitor systems experience a similar reversal. 
In the first case, an ordinary, short-lived
($\sim10^6$ yr) pulsar is created. In the latter, a fast, long-lived
($\sim10^9$ yr) millisecond pulsar should often be the result. It is likely
that type II \bhpsr systems with recycled pulsars would be observable as
radio objects. The small predicted formation rate of such objects, however,
is consistent with having not yet been seen among known pulsars. However,
their formation rate, while 30--60 times smaller than that of ``type 1''
\bhns systems, should mean that they are considerably more common than
black holes with normal pulsar companions, being longer-lived. Also, we
note that the formation rate of ``type 2'' \bhpsr systems is more severely
affected by kick intensity, as strong kicks may widen the system enough to
prevent mass transfer on to the neutron star, leaving an ordinary \bhpsr
system. We conclude that the galactic scale height for such systems should
be substantially smaller than that of ordinary \bhns and \bhpsr systems.
Further details of the \bhpsr systems (including a prediction of orbital
distributions) will be deferred to a future work (Sipior \& Sigurdsson,
in prep). Note finally that the chirp mass of a typical coalescing \bhpsr
system is roughly twice as large as a \nsns binary. As $d_L$ scales with
$M^{5/6}_{chirp}$, this implies at least a five-fold increase in search
volume. Adjusting the estimated $R_6$ rates for this implies that \bhns
coalescence should be seen with substantially greater frequency than
\nsns binaries at all kick intensities, with a conservative maximum of
$7\times10^{-3}$ LIGO I detections per year for low kick strengths. A more
likely ``best bet'' rate is on the order of 2--3$\times10^{-3}$ yr$^{-1}$,
with a mix of high ($\sigma = 450$ \kmsec) and low ($\sigma = 90$ \kmsec)
kick velocities, with 60\% of kicks drawn from the former, the balance from
the latter (as suggested by \citealp{arz01,pfahl:2001}).

\begin{figure}
\epsscale{0.75}
\plotone{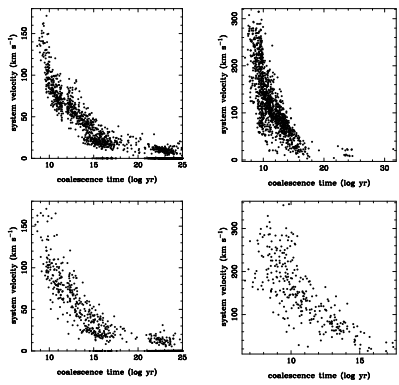}
\figcaption{\label{tvsv}A plot of merger time versus the final velocity of
the \bhbh binaries (left), and \bhns binaries (right). The natal kick
parameters were $\sigma = 90$ \kmsec, $K_{max} = 500$ \kmsec (upper panels),
and $\sigma = 190$ \kmsec, $K_{max} = 500$ \kmsec (lower panels). Of the
systems that remained bound after the second supernova, there is a clear
correlation between increasing system speed and a shortening of the
gravitational radiation merger timescale, from kick-induced high eccentricity
orbits, with enhanced GW radiation at periastron passage.}
\end{figure}

\bhbh systems are shown here to be far more common than either \nsns or
\bhns hybrids. Only a small fraction of these will merge in a Hubble
time, as the larger mass permits the system to remain bound at higher
separations. As well, the greater mass of \bhbh systems implies a more
gradual response to an increasing kick parameter. Indeed, the $R_6$ rate
for bound \bhbh system formation remains high (more than thirty per Myr
in the Milky Way), even when kicks are permitted to reach $10^3$ \kmsec
on a 1.4 M$_{\odot}$ object. Figure \ref{chirp} shows the distribution of
chirp masses for all \bhbh systems that coalesce in less than 10 Gyr. As
can be seen, there is a concentration of chirp masses around 4 M$_{\odot}$,
implying a twentyfive-fold increase in search volume. Allowing for this,
we anticipate a LIGO I detection rate of 0.3--12$\times10^{-3}$ yr$^{-1}$,
with a likely rate of 3--4$\times10^{-3}$ $yr^{-1}$, given the two-component
kick speed distribution described above. Figure \ref{tvsv} shows the strong
correlation between the final system velocity (a function of the kick
magnitude and component masses), and the time required for the system to
coalesce due to emitted gravitational wave energy. A clear dynamic criterion
for coalescence emerges, with slower systems rarely merging in a Hubble
time. This is essentially a selection effect, as the kick speed must be of a
magnitude comparable to the orbital speed of the system to have a dramatic
effect on the merger rate. Relative to the orbital speed, a weak kick will have
no appreciable effect, while a strong kick will disrupt the system. So, the
kick speeds that bring about more rapid mergers are necessarily tuned to the
orbital speed of the system at the second supernova.

\section{Discussion}

We claim a ``best guess'' LIGO I detection rate of 3--4
$\times10^{-3}$yr$^{-1}$ for \bhbh coalescence events, 2--3$\times10^{-3}$
yr$^{-1}$ for \bhns events, and $\sim$10$^{-3}$ \nsns coalescence per year.
LIGO II, with an anticipated order of magnitude increase in sensitivity,
should encompass roughly $10^3$ times the search volume, with a proportional
increase in event detection rates. Unless LIGO I is quite lucky, it seems
unlikely that mergers \emph{from this channel} will be detected until the 
advent of LIGO II in
a few years. With LIGO II, we can anticipate an event rate of many per
year. These are essentially consistent with the most pessimistic detection
rates calculated by \citet{gri01}. There are significant differences
between our results and those in \citet{zwart1}, primarily regarding \bhbh
coalescence, where a negligible merger rate was found. This discrepancy is
likely due to our lower cutoff for black hole formation, where we adopt
a value of 20 M$_{\odot}$, against the 40 M$_{\odot}$ cutoff adopted by
the latter. While recent models of \citet{fryer1999} have established a
lower black hole mass cutoff, more work is needed to resolve this area
of contention. LIGO-measured coalescence rates should provide a clearer
resolution on this point.

It is instructive to compare our results with those obtained in two recent
population synthesis studies. \citet{fryer:1999} performed detailed synthesis
calculations in the context of black hole-accretion driven gamma ray bursts.
As a result, \bhbh binaries are not considered; however, \bhns and \nsns
systems are. For the Salpeter IMF, and a Gaussian kick distribution peaked
at 100 \kmsec, FWHM of 50 \kmsec (roughly corresponding to our $\sigma =
90$ \kmsec kicks), the $R_6$ rate of \bhns formation is identical ($R_6
= 16$). We predict many more \nsns ($R_6$ of 11, as opposed to 4.2).
This discrepancy arises from the inclusion of hypercritical accretion in
\citeauthor{fryer:1999} scenario I for \nsns formation. The result is that all
neutron stars that pass through a common envelope phase become black holes,
removing a significant channel of \nsns pair formation. At higher kick
intensities, the \bhns formation rates remain in good agreement, whilst the
\nsns rates become more disparate. This is likely because the two high speed
kick distributions are not the same, and the discrepancy affects \bhns systems
less dramatically because of the larger mass involved.

Very recently, the paper of \citet{bel:2001} performed a similar
comprehensive study, considering a wide variety of kick paradigms and
population synthesis models. Our anticipated LIGO I detection rates are
consistently an order of magnitude lower than those reported by the authors for
their ``standard model''. Though the kick distribution chosen for this standard
model is more intense than in our own work (two Maxwellian distributions, with
$\sigma$ = 175 \kmsec and $\sigma$ = 700 \kmsec, with 80\% of kicks drawn from
the former, and the balance from the latter distribution), the mean neutron
star mass is allowed to be much higher (up to 3 M$_{\odot}$, allowing more 
systems to remain bound, \emph{and} shed gravitational radiation more quickly.
Thus, the overall effect is to enhance the anticipated \nsns and \bhns merger
rates. It should be noted that our calculated merger rates fall within the
range of merger rates calculated by \citeauthor{bel:2001} when all posited
model assumptions are considered. The difference between our results and the
standard model of the authors for \bhbh systems is larger, by roughly two
orders of magnitude (though our results still fall within the wider range
resulting from considering all the models presented). Our model assumptions are
greatly discrepant at one point in particular; namely, the assumed mass
function of newly-formed black holes as a function of ZAMS mass. Where we take
a rough interpolation between model results presented in \citet{fryer1999}, the
authors use a much more sophisticated relationship between the ZAMS mass and
the initial mass of the resulting black hole. This mass function tends to
produce substantially more massive holes than those generated from our
interpolation, and we believe this to be the principal cause of the 
discrepancy in merger rates for \bhbh binaries.

After the neutron star/black hole mass cutoff, the second pivotal assumption
concerns the distribution of natal kick speeds, and indeed, whether black
holes can receive kicks under any circumstances. The cases of Nova Sco
1994 and XTE J1118+480 show that our natal kick assumptions are plausible.
However, we suffer from a paucity of data concerning the frequency and
scaling (with respect to mass) of black hole kicks. There is mounting
evidence that neutron star kick distributions are bi-modal \citep{arz01},
allowing many more \nsns systems to survive and merge after the second
supernova. The greater mass of black holes means that uncertainties in
these distributions have less effect on the coalescence of \bhns and \bhbh
systems, as can be seen in table \ref{r6rates}. Coupling this fact with the
``residual'' merger rate one gets when black holes are not allowed natal
kicks at all (cf. table \ref{r6nobhkicks}), we argue that the effects of
uncertainty in the natal kick models will not affect the predicted $R_6$
rates of \bhbh and \bhns mergers by more than factors of a few. Interestingly,
the large number of \bhbh binaries which remain bound suggests that another
merger channel; namely, a merger due to hardening resulting from
external dynamical perturbations \citep{sig1}. We do not quantify this effect
here, but note that the extent and duration of a star-formation episode could
have a notable effect on the rate at which such interactions occur.

Black holes experience natal kicks only when formation of the event horizon
is delayed, from fallback material and/or initial spin support. Therefore
it is reasonable to posit a correlation between natal kicks and a high
initial spin for the nascent black hole. In a \bhbh system, the first
hole is gradually spun down by interaction with the stellar wind of its
companion, and will likely be a slow rotator even if it began with maximal
spin. The second hole will not experience this, and should retain its high
spin. We therefore predict that, for BH systems that experience natal kicks,
LIGO will see a slow rotating primary, with a (generally less massive)
black hole companion with near maximal spin. If the components do not
experience natal kicks, we would anticipate low spins all around, with
the coalescence rate primarily determined by the wind mass loss rate and
the boundary criterion for black hole formation. This spin signature is
a directly testable hypothesis---LIGO II should provide a large array of
high-quality sources to verify this prediction, and allow for careful
testing of gravitational wave phenomenology.

A similar spin correlation may be found in \bhpsr systems, with the two
categories described above having different anticipated signatures. The type
I systems (black hole first, followed by standard pulsar) should have two
slow rotators, as no pulsar recycling is possible. Type II systems (neutron
star forms first, is spun up by black hole progenitor, black hole forms)
should have at least one rapid rotator, and possibly two, if the black hole
is spun up during its formation, as above. These spin correlations may
be demonstrated with high-S/N data from LIGO II. More will be said about
the spin signatures of \bhpsr systems in an upcoming paper (Sipior \&\
Sigurdsson, in prep), in addition to a study of the orbital parameters of
such systems, and the ratio of \bhpsr to NS-PSR systems, which is relevant
for new tests of General Relativity, should these systems be detected.

Model uncertainties, especially involving the details of black hole
formation, still dominate the estimates of coalescence rates to be seen
by LIGO. Nonetheless, it is becoming clear that \bhbh mergers form a
significant channel, comparable to the rates anticipated for \nsns systems.
While LIGO I detection rates appear too small to catch any merger events from
this channel, we may still be lucky enough (at the level of a few percent) to
serendipitously catch events from this channel (especially given incremental
upgrades in the sensitivity of LIGO I throughout its operational lifetime).
LIGO II should easily see a large number of these events every year.

\begin{deluxetable}{c*{9}{c}}
\tabletypesize{\small}
\tablecomments{$R_6$ formation and merger rates for \bhbh,
\bhns, \nsns and \bhpsr systems. $\alpha$ is the index of the IMF power-law, $\sigma$
provides the dispersion of the Maxwellian distribution of kick speeds, and
the kick cutoff gives the maximum permitted kick speed (acting on a 1.4
M$_{\odot}$ neutron star). Black hole kicks are permitted below a ZAMS mass of
40 M$_{\odot}$, and are scaled to the compact object mass. The \bhpsr column
refers to a subset of the \bhns systems, where the pulsar is recycled by mass
transfer from the black hole progenitor star. The mass transfer should in many
cases spin-up the pulsar to millisecond-order periods.
\label{r6rates}}
\tablehead{\colhead{$\alpha$} & \colhead{$\sigma$ (\kmsec)} & \colhead{BH-BH} &
\colhead{$\tau<\tau_h$} & \colhead{BH-NS} & \colhead{$\tau<\tau_h$} &
\colhead{NS-NS} & \colhead{$\tau<\tau_h$} & \colhead{BH-PSR$_2$} &
\colhead{$\tau<\tau_h$}}

\startdata
-2.0 & 90 & 80 & 1.6 & 26 & 8.2 & 11 & 8.4 & 0.62 & 0.44\\
-2.0 & 190 & 68 & 0.59 & 5.9 & 2.5 & 1.7 & 1.5 & 0.11 & 0.09\\
-2.0 & 450 & 62 & 0.11 & 0.38 & 0.2 & 0.09 & 0.06 & 0.0 & 0.0\\
-2.35 & 90 & 48 & 1.2 & 16 & 5.5 & 11 & 7.9 & 0.47 & 0.37\\
-2.35 & 190 & 41 & 0.55 & 4.0 & 1.7 & 1.6 & 1.3 & 0.07 & 0.07\\
-2.35 & 450 & 35 & 0.04 & 0.2 & 0.09 & 0.07 & 0.05 & 0.0 & 0.0\\
\enddata
\end{deluxetable}

\begin{deluxetable}{c*{9}{c}}
\tabletypesize{\small}
\tablecomments{$R_6$ formation and merger rates. Identical to table
\ref{r6rates}, save that black hole kicks are not permitted under any
circumstances. This clearly elevates the number of \bhbh and \bhns
systems seen. Interestingly, the application of natal kicks increases the
\emph{fractional} merger rate in \bhbh systems, though the large number of
\bhbh systems which remain bound when kicks are not present means that the
number of merging systems is still comparable, especially for the most
energetic kick parameters.
\label{r6nobhkicks}}
\tablehead{\colhead{$\alpha$} & \colhead{$\sigma$ (\kmsec)} & \colhead{BH-BH} &
\colhead{$\tau<\tau_h$} & \colhead{BH-NS} & \colhead{$\tau<\tau_h$} &
\colhead{NS-NS} & \colhead{$\tau<\tau_h$} & \colhead{BH-PSR$_2$} &
\colhead{$\tau<\tau_h$}}

\startdata
-2.0 & 90 & 250 & 0.81 & 31 & 9.0 & 10 & 7.8 & 0.63 & 0.53\\
-2.0 & 190 & 240 & 0.62 & 12 & 5.5 & 1.8 & 1.5 & 0.08 & 0.06\\
-2.0 & 450 & 240 & 0.77 & 2.0 & 1.1 & 0.04 & 0.03 & 0.0 & 0.0\\
-2.35 & 90 & 170 & 0.55 & 21 & 6.5 & 11 & 8.1 & 0.55 & 0.47\\
-2.35 & 190 & 170 & 0.71 & 7.8 & 3.4 & 1.7 & 1.4 & 0.07 & 0.07\\
-2.35 & 450 & 170 & 0.32 & 1.4 & 0.83 & 0.04 & 0.04 & 0.0 & 0.0\\
\enddata
\end{deluxetable}

\clearpage

\clearpage

\clearpage

\clearpage

\clearpage

{\it Acknowledgments:\/}
Both authors would like to profusely thank Onno Pols and Joshua Bloom for
access to the code base that we built upon here.
Steinn Sigurdsson would like to acknowledge the support of the Center for
Gravitational Wave Physics, and the hospitality of the Aspen Center for
Physics. The Center for Gravitational Wave Physics is supported by the NSF
under co-operative agreement PHY 01-14375.
Michael Sipior is supported in part by NASA through grant GO01152 A,B from the
Smithsonian Astrophysical Observatory. He would like to acknowledge travel
support kindly provided by the Zaccheus Daniel Foundation.

%%%%%%%%%%%%%%%
% Reference List
%%%%%%%%%%%%%%%

\end{document}